# Molecular Dynamics Simulation of Methanol-Water Mixture

**Palazzo Mancini, Mara Cantoni**

University of Urbino Carlo Bo

**Abstract**

*In this study some properties of the methanol-water mixture such as diffusivity, density, viscosity, and hydrogen bonding were calculated at different temperatures and atmospheric pressure using molecular dynamics simulations (MDS). The results were compared with the available experimental data as well as some theoretical models; overall indicating a good agreement. This shows the useful and effective application of MDS for determination of physical properties.*

## 1- Introduction

As molecular simulations are considered useful tools for the interpretation of experimental data, two methods namely Monte Carlo (MC) and Molecular Dynamics (MD), can be applied. This allows us to determine macroscopic properties by evaluating exactly a theoretical model of molecular behavior using a computer program. [1,2]

Molecular simulation considers small size systems, at a typical scale of a few nanometers, and determines their behaviour from a careful computation of the interactions between their components. These computations take much more computer time than classical thermodynamic models (i. e. from a few hours to several weeks of computing time, depending on system size). As such, molecular simulation is now considered in the chemical industry as capable of filling the existing gap between experimental data and engineering models in various circumstances such as unknown chemicals, extreme conditions of temperature or pressure, and toxic compounds.

The first part of this article provides a brief description of Molecular dynamic simulation and how it can be analyzed for approving experimental data. In the second part, different properties of the system will be calculated by molecular dynamic simulation. Results will then be compared with available experimental data and theoretical models.

## 2- Molecular dynamics method
### 2.1- What is molecular dynamics method?
Molecular dynamics simulation involves integration of Newton's equation of motion for each atom in an ensemble in order to generate information on the variation in the position, velocity, and acceleration of each particle as a function of time. The equations solved are [3]:

$$f_i = m_i a_i \qquad (1)$$

$$f_i = -\frac{\partial U}{\partial r_i} \qquad (2)$$

Where $f_i$ is the force exerted on the $i^{th}$ atom, $m_i$ is its mass, and $a_i$ is its acceleration. The force acting on each atom is obtained from a potential function $U$, such as that depicted below [3]:

$$U = \sum \frac{k_{ij}^b}{2}(r^{ij} - r^{ij}_{eq})^2 + \sum_{angle} \frac{k_{ijk}^\theta}{2}(\theta^{ijk} - \theta^{ijk}_{eq})^2 + \sum_{dihedrals} k^\phi \left[1+\cos(n(\phi - \phi^{eq}))\right] + \sum_{i<j} \left[\frac{A_{ij}}{r_{ij}^{12}} - \frac{B_{ij}}{r_{ij}^{6}} + \frac{q_i q_j}{4\pi\varepsilon_0 r_{ij}}\right] \qquad (3)$$

Where the first three terms reflect intramolecular interactions between covalently bonded atoms and the last term describes the non-bonded interactions (van der Waals and electrostatic). The constants $k^b$, $k^\theta$, and $k^\phi$ are force constants for the covalent bonds, bond angles, and dihedrals, respectively, $r_{ij}$ are the distances between atoms i and j, $A_{ij}$ and $B_{ij}$ are the Lenard Jones parameters, and $q$ is a partial charge [3].

To calculate the properties of a real system, it is convenient to define an ensemble. An ensemble can be regarded as an imaginary collection of a very large number of systems in different quantum states with common macroscopic attributes. The ensemble average of any dynamic property (A) can be obtained from the relationship:

$$\langle A \rangle = \sum_i A_i p_i \qquad (4)$$

Where $A_i$ is the value of $A$ in quantum state i, $p_i$ represents the probability of observing the i-th state, and the angled brackets denote an ensemble average. The time averaged properties of the real system are related to the ensemble average by invoking the following assumption:

$$A_{t=\infty} = \langle A \rangle \qquad (5)$$

The above equivalence of the time – average and ensemble – average is called the Ergodic hypothesis. Form of $p_i$ is determined by the macroscopic properties that are common to all of the systems of the ensemble.

### 2.2- Calculation of physical properties
Prediction of transport properties of fluids of industrial interest is one of the main motivations to perform MD simulations [4].

As it is not possible to observe individual atoms or molecules directly, various models are used to describe and/or predict the properties of a system with molecular dynamic methods. On the other hand, it is possible to calculate macroscopic properties of the mixtures by calculating the interaction of the molecules. The partial density is one macroscopic property that could be calculated using molecular dynamics simulation. Also transport properties like the diffusion coefficient and shear viscosity can be computed directly from an MD simulation too [5]. In an equilibrium molecular dynamics (EMD) simulation, the self-diffusion coefficient $D^S_\alpha$ of component α is computed by taking the slope of the mean square displacement (MSD) at a long period of time:

$$D^S_\alpha = \frac{1}{2dN_\alpha} \lim_{t \to \infty} \left\langle \sum_{i=1}^{N_\alpha} (r^\alpha_i(t) - r^\alpha_i(0))^2 \right\rangle \qquad (6)$$

Where $N_\alpha$ is the number of molecules of component α, $d$ is the spatial dimension of the system, $t$ is the time, and $r^\alpha_i$ is the center of mass of molecule i of component α. Equivalently, $D^S_\alpha$ is given by the time integral of the velocity autocorrelation function:

$$D^S_\alpha = \frac{1}{dN_\alpha} \int_0^\infty \left\langle \sum_{i=1}^{N_\alpha} v^\alpha_i(t) v^\alpha_i(0) \right\rangle \qquad (7)$$

Where $v^\alpha_i$ is the center of mass –velocity of molecule i of component α. Equation (6) is known as the Einstein equation and equation (7) is often referred to as the Green-Kubo relation. [6]

Also shear viscosity can be calculated using the following equation:

$$\eta = \frac{V}{k_B T} \int_0^\infty \left\langle \sigma_{xy}(t) \sigma_{xy}(0) \right\rangle dt \qquad (8)$$

Where $\sigma_{xy}$ is the pressure tensor of the components. $k_B$ is the Boltzman constant and T is the temperature of the system.

### 2.3- GROMACS Software

GROMACS is an acronym for GROningen MAchine for Chemical Simulation. It is the latest release of a versatile and very well known optimized package for molecular simulation. The package is a collection of programs and libraries for the simulation of molecular dynamics and the subsequent analysis of trajectory data. Although it is primarily targeted at biological molecules with complex bonded interactions, the very effective implementation of nonbonded force calculations makes GROMACS suitable for all kinds of molecular dynamics simulations based on pair potentials. Apart from normal potential functions like Lenard-Jones, Buckingham and Coulomb, it is possible to use arbitrary forms of interactions with spline-interpolated tables [7]. The GROMACS molecular dynamics code contains several algorithms that have been developed in the Berendsen group. The leap-frog integrator is used in two versions, for molecular dynamics and stochastic dynamics [8].

### 3. Result and Discussion
### 3.1-System definition
In Gromacs software, the system is defined by its size and shape, the number and types of molecules it contains, and the coordinates and velocities of all atoms. Velocities may also be generated from a Maxwellian distribution [9].

In this study MD simulation was used to examine some physical and transport properties of the mixture of methanol ñ water. This system consists of 216 molecules of water and 216 molecules of methanol. Simulations are carried out using a fully atomistic system with the maximum size 4.6 * 2.3 *2.3 nm. The box is shown in Fig. 1. Gromacs software was used for the MD calculations. Simulation was performed using GROMOS 96 [10, 11] force field for 200 ps. MD simulations are performed at atmospheric pressure with four different temperatures 273, 283, 288, 293 k at the isothermal-isobaric ensemble (NPT). Table 1 shows the Lenard-Jones parameters and partial charge for methanol molecule. Water molecules are modeled as SPC water model with their related parameters [10].

### 3.2- Diffusion coefficient of the methanol in the mixture

As the first step, diffusion coefficient of methanol molecules in a mixture of methanol-water has been studied by MD method. For this reason mean square displacement (MSD) of methanol molecules in the mixture was calculated by Gromacs software. Fig. 2 shows MSD values versus time for methanol molecules diffusing in the methanol-water mixture at four temperatures 273, 283, 288 and 293 k. These graphs show that the behavior of MSD versus time for $0<t<200$ ps is linear, so Einstein equation can describe the diffusion behavior of methanol. Calculated Diffusion coefficients of methanol in the mixture by Einstein equation are shown in Table 1 and Fig. 3.

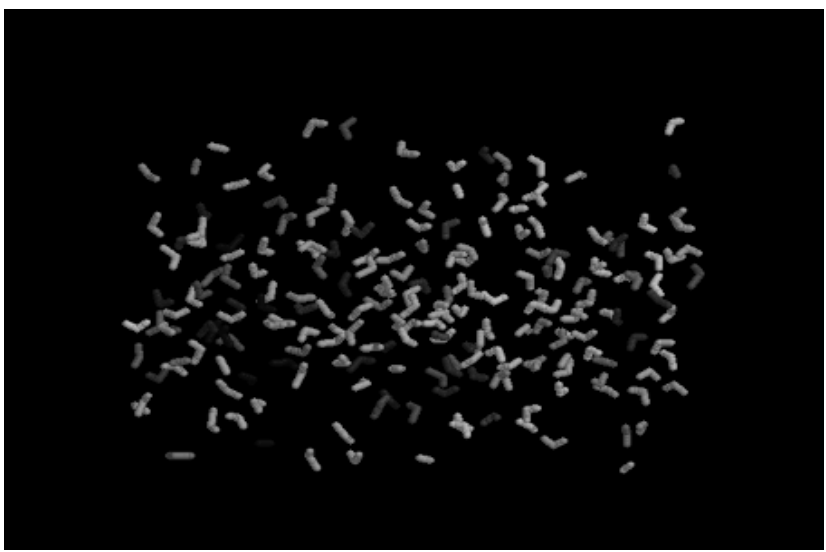

**Figure 1.** Computer generated image of a methanol-water box

**Table 1.** The partial charge and Lenard-Jones parameters for methanol molecule [10]

| Atom type | Charge | Lenard Jones parameters | |
|---|---|---|---|
| | | C6 | C12 |
| CH3 | 0.29 | 8.8758e-3 | 17.8426e-6 |
| O | -0.69 | 2.6169e-3 | 2.5231e-6 |
| H | 0.4 | 0 | 0 |

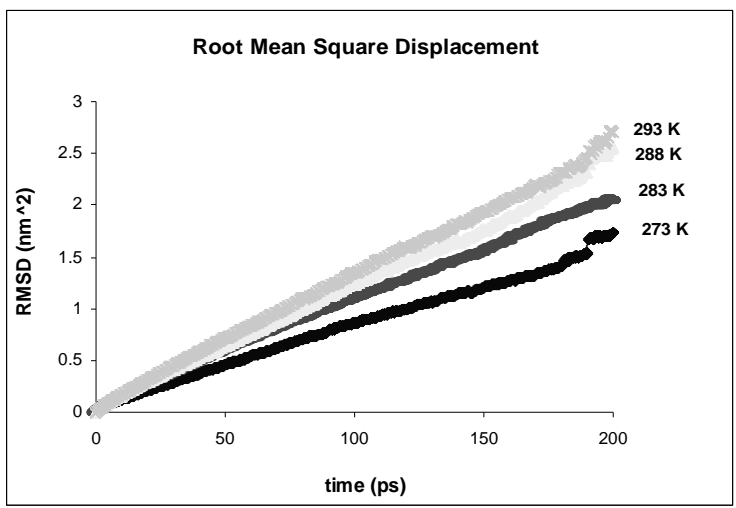

**Figure 2.** Root Mean square displacement of methanol molecules at different temperatures

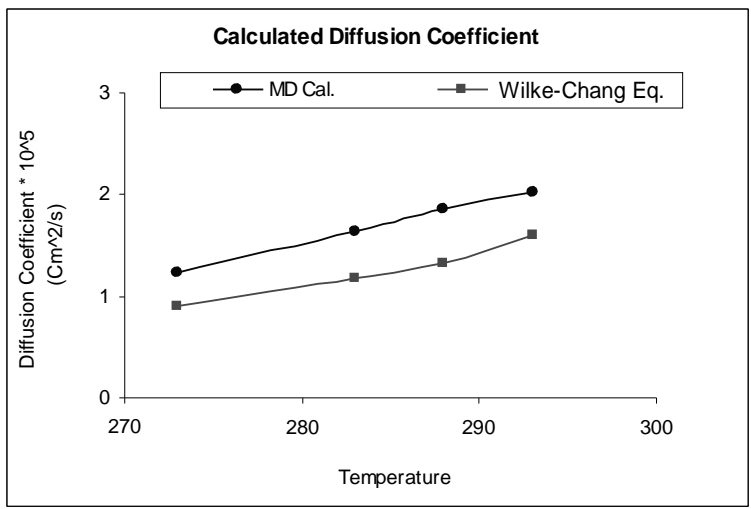

**Figure 3.** The effect of temperature on diffusion coefficient.

To evaluate the calculated results, the diffusion coefficient of this mixture was calculated by Wilke- Chang equation [12]. Results are shown in Table 1 and Fig 3:

$$D_{AB} = \frac{(117.3 \times 10^{-18})(\varphi M_B)^{1/2} T}{\mu V_A^{0.6}} \quad (9)$$

The available experimental data for diffusion coefficient of methanol in water are shown in Table 2 [13]. The averaged error between MD results and experimental value is about 12%, while this is about 10% between experimental value and Eq.9.

### 3.3- Calculation of density

In the second step, the density of methanol-water mixture was calculated at different temperatures and different points in the box. The results are shown in Table 3 and Fig. 4 and 5. The comparison between experimental [13] and calculated results shows a good agreement between these data and the averaged error is less than 1%.

**Table 2.** Calculated Diffusion Coefficient of Methanol in Water by MD method

| Temperature (K) | Diffusion coefficient by MD (cm$^2$/s) | Diffusion coefficient by Eq. 9 (cm$^2$/s) | Experimental value [11] |
|---|---|---|---|
| 273 | 1.23 * 10$^{-5}$ | 0.98 * 10$^{-5}$ | - |
| 283 | 1.63 * 10$^{-5}$ | 1.18* 10$^{-5}$ | - |
| 288 | 1.80*10$^{-5}$ | 1.42* 10$^{-5}$ | 1.56 * 10$^{-5}$ |
| 293 | 2.03*10$^{-5}$ | 1.60* 10$^{-5}$ | 1.80 *10$^{-5}$ |

**Table 3.** Calculated and experimental density in different temperatures

| Temperature (K) | Cal. Density (kg/m^3) | Exp. Density (kg/m^3) [11] | Error % |
|---|---|---|---|
| 273 | 936.10 | 928.70 | 0.8 |
| 283 | 927.27 | 922.10 | 0.6 |
| 288 | 918.15 | 918.50 | 0.04 |
| 293 | 915.54 | 915.60 | 0.006 |

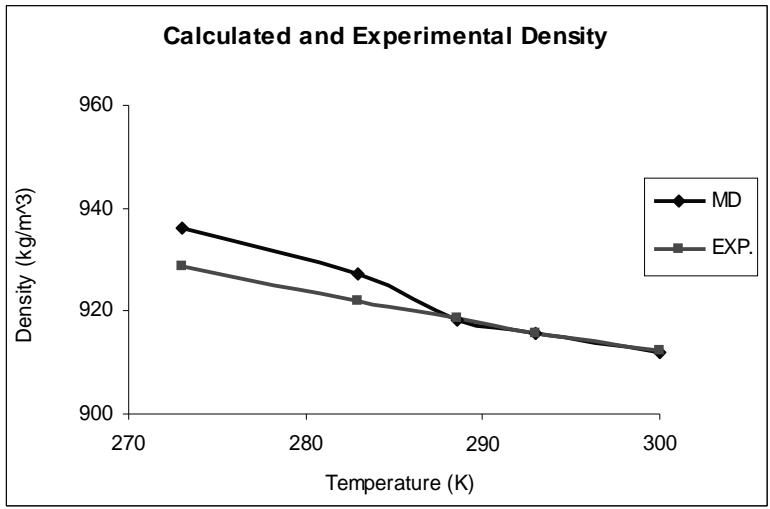

**Figure 4.** The effect of temperature on density for methanol-water mixture

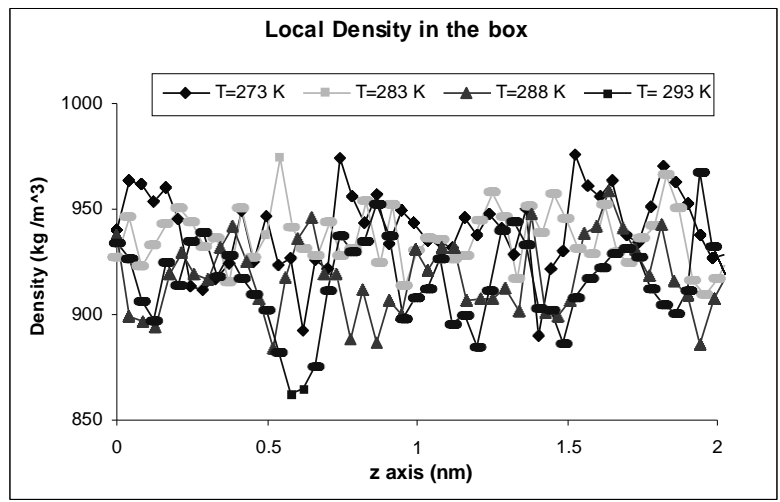

**Figure 5.** Calculated density in different points in the box by MD method

### 3.4- Shear viscosity

One of the important properties of the mixtures is viscosity. It is possible to calculate this macroscopic property of the mixture by calculating the interaction of the molecules in microscopic scale. Here the viscosity of the methanol-water mixture was calculated by MD simulation and the results are compared with experimental data.

As it was shown in Table 4 and Fig. 7, viscosity is decreased by increasing temperature, the same as experimental value.

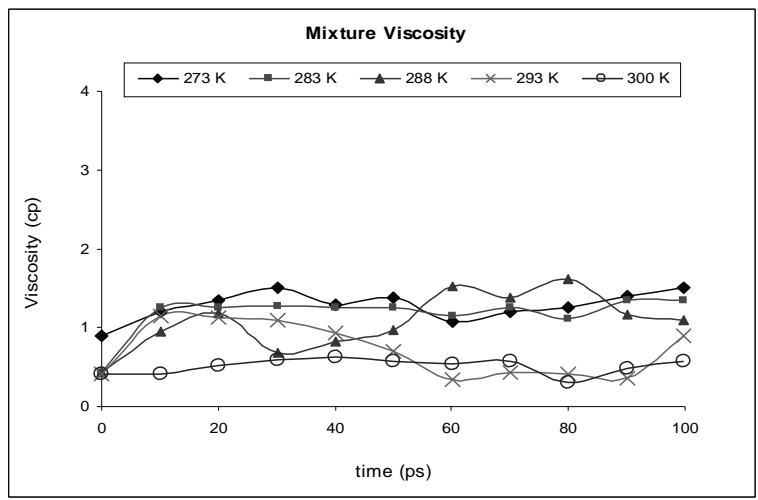

**Figure 6.** Viscosity of the methanol-water mixture at different temperatures

**Table 4.** Average calculated and experimental viscosity of the methanol-water mixture

| Temperature | 273 K | 283 K | 288 K | 293 K | 300 K |
|---|---|---|---|---|---|
| **MD Calculated value** | 1.27 | 1.07 | 0.95 | 0.71 | 0.51 |
| **Experimental value [12]** | 1.06 | 0.9 | 0.85 | 0.74 | 0.6 |
| **Error %** | 19.8 | 18.0 | 11.7 | 4.1 | 7 |

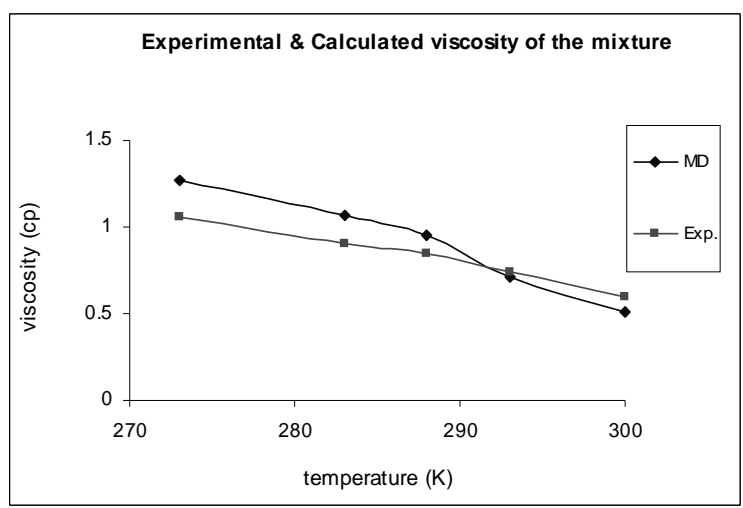

**Figure 7.** The effect of temperature on viscosity for methanol-water mixture

### 3.5- Hydrogen Bond

Many thermodynamical properties (e.g., melting and boiling point) of the mixture depend on the strength of the hydrogen bonds (H-bond) formed among molecules. The H-bond is a special attractive interaction parameter that exists between an electronegative atom and a hydrogen atom bonded to another electronegative atom. Since water has two hydrogen atoms and one oxygen atom, it can form an H-bond between the Oxygen and the H atoms. Each water molecule can form up to four hydrogen bonds at the same time (two through its two lone pairs, and two through its two hydrogen atoms). The average number of H-bonds per molecule calculated in bulk water varies approximately from 2.3 to 3.8 according to the water model and the way used to define the H-bond. [14, 15] In this study the number of hydrogen atoms between water molecules and water-methanol molecules were calculated. The results, shown in Figs. 8 and 9, indicate that the number of hydrogen bonds between water molecules is from 2 to 3 but it decreases for water and methanol molecules down to 1. The hydrogen bond distribution is also plotted in the box, where the most probable ones form near the center, Fig 10.

### 4. Conclusion

In this paper molecular dynamic theoretical method is presented to calculate density, shear viscosity, hydrogen bonding and diffusion coefficient for the mixture of methanol-water. Gromacs software was used for this purpose. The simulation was performed up to 200 ps. The results in Table 1 and Fig. 3 clearly show that the calculated diffusion coefficient by MD method indicate a good agreement and an almost similar trend with those obtained from the Wilke-Chang equation, both of which are approved by experimental data. These values, however increase linearly with temperature (D~T).

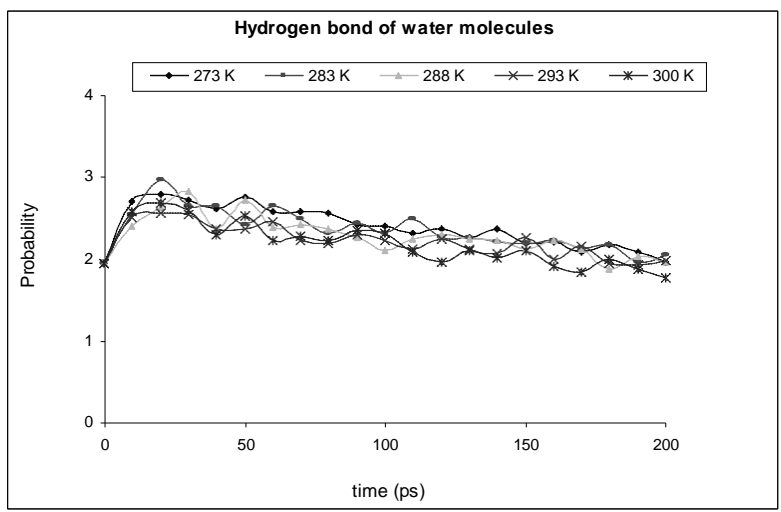

**Figure 8.** Hydrogen bond of water molecules in the mixture

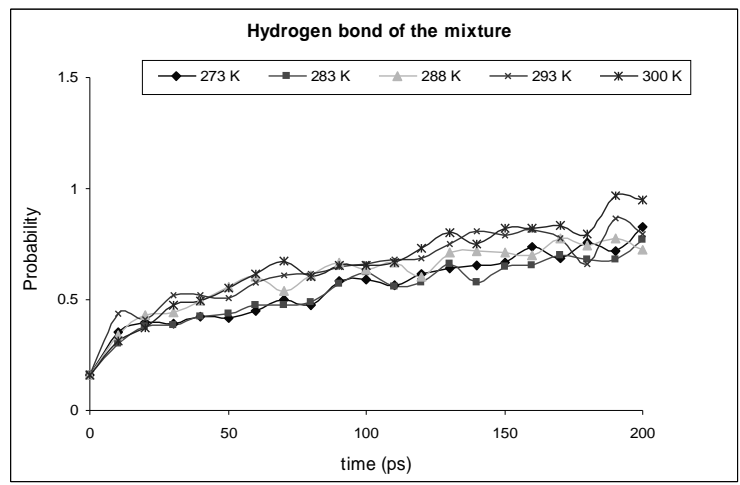

**Figure 9.** Hydrogen bond of water and methanol molecules in the mixture

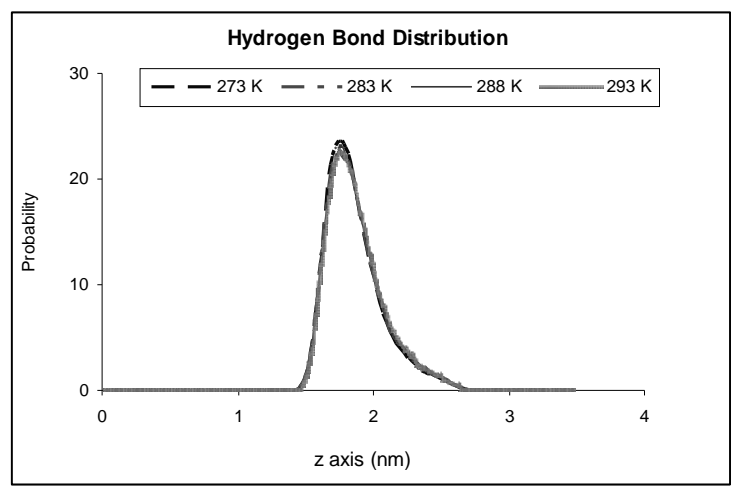

**Figure 10.** Hydrogen bond distribution of methanol-water mixture in the box

Density and viscosity similar to that of diffusion coefficient are affected by temperature. The density decreases with temperature with a decreasing deviation between MD and experimental data. This shows that MD is much more reliable at higher temperatures. In the case of viscosity, it also decreases with temperature with a decreasing deviation. After that, it decreases but with a lower deviation of MD with those of the experimental data. The reference temperature for MD calculation in the input files of the simulation is 300 K, so some of the calculated properties such as viscosity and density at the points near this temperature have the minimum deviation from the experimental data.

The MD calculation for pure water in this study estimated H-bond numbering between 2 to 3 which is in a good agreement with the previous studies [14-18]. This number decreases considerably for the case of water-methanol molecules. This is due to the lower polarity of methanol molecules, compared to that of water molecules. Hydrogen bond formation has the most distribution probability at the center of the box and the temperature has no effect on distribution. In the center of the box, there are water and methanol molecules in all the directions, so the hydrogen bond has the maximum distribution in this part of the system, but in the points near the sidewall of the system the number of molecules decreases so the hydrogen bonding has the minimum distribution in these points.

Overall, the results show the reliability of the MD method to calculate the physical properties of the systems. One can conclude that, this method may be used to determine the macroscopic properties of the systems, instead of practical methods in the laboratory.

**Nomenclature**

T      Temperature (K)
$M_B$      Solvent molecular weight (kg/kmol)
$\varphi$      Solvent accumulation coefficient (-)
$\mu$      Viscosity (kg/m.s)
$V_A$      solute molar volume at boiling point ($m^3$/kmol)